\documentclass{bioinfo}
\copyrightyear{2012}
\pubyear{2012}
\usepackage[multiple]{footmisc}

\begin{document}
\firstpage{1}

\title[Biospectrogram: a tool for spectral analysis of biological sequences]{Biospectrogram: a tool for spectral analysis of biological sequences}
\author[Sample \textit{et~al}]{ Nilay Chheda\,$^{1,\footnote{These authors contributed equally to the project and should be considered co-first 
authors}}$, Naman Turakhia \,$^{1,\footnotemark[1]}$, Manish K. Gupta \,$^1$\footnote{to whom correspondence should be addressed}, Ruchin Shah\,$^{1}$ and  Jigar Raisinghani\,$^{1}$}
\address{$^{1}$ Laboratory of Natural Information Processing, 
Dhirubhai Ambani Institute of Information and Communication Technology, Gandhinagar, Gujarat,  382007 India.
}
\history{Received on XXXXX; revised on XXXXX; accepted on XXXXX}

\editor{Associate Editor: XXXXXXX}

\maketitle

\begin{abstract}
\section{Summary:} Biospectrogam is an open-source software for the spectral analysis of DNA and protein sequences. The software can fetch (from NCBI server), import and manage biological data. One can analyze the data using Digital Signal Processing (DSP) techniques since the software allows the user to convert the symbolic data into numerical data using $23$ popular encodings and then apply popular transformations such as Fast Fourier Transform (FFT) etc. and export  it. The ability of exporting (both encoding files and transform files) as a MATLAB\textsuperscript{\textregistered} .m file gives the user an option to apply variety of techniques of DSP. User can also do window analysis (both sliding in forward and backward directions and stagnant) with different size windows and search for meaningful spectral pattern with the help of  exported  MATLAB\textsuperscript{\textregistered} file in a dynamic manner by choosing time delay in the plot using Biospectrogram.  Random encodings and user choice encoding  allows software to search for many possibilities in spectral space. 
\section{Availability:}
Biospectrogam is written in Java\textsuperscript{\textregistered} and is available to download freely from http://www.guptalab.org/biospectrogram. Software has been optimized to run 
on Windows, Mac OSX and Linux. User manual and you-tube (product demo) tutorial is also available on the website.  We are in the process of acquiring  open source license for it.
\section{Contact:} \href{mankg@computer.org}{mankg@computer.org}
\end{abstract}

\section{Introduction}
Molecular biology has shown tremendous progress  in the last decade because of various genome projects producing vast amount of biological data.  This has resulted in Encode project (http://encodeproject.org) that classifies all the basic DNA elements of Human genome. This also gives us new insight into numerous molecular mechanism. In order to understand  the digital biological data, people use different techniques from mathematics, computer science, etc.  Digital signal processing (DSP) is a fundamental concept in information and communication technology (ICT).  A natural question arises ``Can DSP techniques  help us to understand the digital biology?" It turns out that the DSP techniques are playing a major role in biology and have given  birth to a new branch called genomic signal processing \citep{shmulevich2007genomic}. To analyse the genomic data, researchers first convert the symbolic data (example DNA or protein data) into numerical data by applying a suitable map \citep{Kwan,Arniker2012} and then by applying signal processing transforms such as Fourier etc. to study the desired biological properties \citep{citeulike:3895919}. In this work, we present a tool, Biospectrogram, which can help researchers to apply different encodings on the biological data and apply certain transformations to do the spectral analysis. User can also export the files (encoded or transformed) to popular  MATLAB\textsuperscript{\textregistered} software \citep{MATLAB:2010} to do the direct analysis.
\section{Implementation and Features}
The tool Biospectrogram has $4$ major components viz. data collector, encode, transforms and export $\&$ plot.  One can use the tool in DNA or Protein mode by using the switch button. The tool has two main windows viz display window for displaying the data (collected or encoded data) and work window (encoded or transform data) to show the work.  Data collector module provides a direct fetching of DNA data (both fasta and genebank file formats) from  National Center of Biotechnology Information (NCBI) server by taking accession number from user which can be encoded using encode button. User can also import the files from his own machine/network.  One can also select a portion of the data from the window and do further processing. 
One popular encoding map is the Voss representation \citep{1195219} which maps the nucleotides $A, C, G,$ and $T$ from DNA space into the four binary indicator sequences $x_A[n]$, $x_C[n]$, $x_G[n]$, and $x_T[n]$ showing the presence (e.g. $1$) or absence (e.g. $0$) of the respective nucleotides. Similar indicator maps are available for protein space.  
\begin{figure*}[!tpb]
\centerline{\includegraphics[scale=0.46]{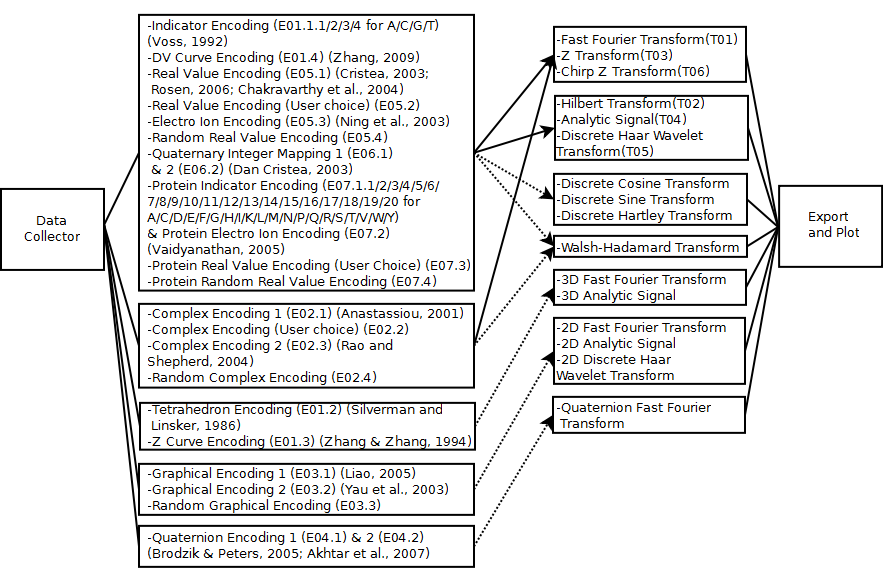}}
\caption{Basic architecture of Biospectrogram showing $4$ major components (data collector, encode, transforms and export $\&$ plot).  Different relationships between $23$ encodings and transformations (with solid arrows possible in our tool)  and others possible broken arrows using third party software MATLAB.}\label{fig:02}
\end{figure*}
Different  possible encodings ($23$ available in our tool) and transformations ($6$ available in our tool) are shown in Figure  ~\ref{fig:02}.  While applying encoding  user has to select the fetched file from the first dropdown list and encoding scheme from the second dropdown list.  The fasta file of the DNA sequence is shown in the display window and the encoded output is shown in the work window.  After encoding the fetched DNA sequence or protein sequence, one can apply suitable transforms (see  Figure  ~\ref{fig:02}) available in the tool. To apply other transforms (not available in our tool) and filters etc. one can export the encoded files to MATLAB\textsuperscript{\textregistered} .m files and do the further analysis.  For exhaustive search of a pattern, a window analysis can be done with our tool. The window button allows the user to set a window size while moving the window in both directions (forward and backward) using sliding window option whereas using stagnant window option user can select a portion of the sequence for the power spectrum from all its indicator sequences. By choosing appropriate delay time in the preferences, one can plot the transformation's output of our tool by exporting the transformation files to MATLAB\textsuperscript{\textregistered} .m files and observe the signal in a smooth automatic manner with a delay of time set by user. 
\nocite{1195219,SilvermanLinsker,zhangzhang,citeulike:4180094,939833,1346354,liao,Yau2003,citeulike:6778043,4365821,cristea, rosen, chakravarthy,1227391,DanCristea:2003:LSF:774474.774488,Vaidyanathan05genomicsand}
\bibliographystyle{natbib}
\bibliography{dspgen}
\end{document}